\documentclass[12pt]{article}
\usepackage{amsmath,amsfonts,epsf}

\renewcommand{\Im}{\operatorname{Im}}
\newcommand{\Tr}{\operatorname{Tr}}
\newcommand{\dd}{\text{d}}
\newcommand{\q}{\boldsymbol q}
\newcommand{\p}{\boldsymbol p}

\newcommand{\I}{\boldsymbol I}
\newcommand{\pphi}{\boldsymbol \phi}

\begin{document}

\noindent
mpi-pks/9907008
~\vspace{3.0cm}

\centerline{\LARGE Geometrical theory of diffraction and spectral statistics}
\vspace{2.0cm}
\centerline{\large Martin Sieber\footnote{E-mail: 
sieber@mpipks-dresden.mpg.de}}
\vspace{0.5cm}

\centerline{
Max-Planck-Institut f\"ur Physik komplexer Systeme,
N\"othnitzer Str.\ 38, 01187 Dresden, Germany}

\vspace{5.0cm}
\centerline{\bf Abstract}
\vspace{0.5cm}
We investigate the influence of diffraction on the statistics
of energy levels in quantum systems with a chaotic classical
limit. By applying the geometrical theory of diffraction we
show that diffraction on singularities of the potential can
lead to modifications in semiclassical approximations for
spectral statistics that persist in the semiclassical
limit $\hbar \rightarrow 0$. This result is obtained by
deriving a classical sum rule for trajectories that connect
two points in coordinate space.

\vspace{2.5cm}

\noindent PACS numbers: \\
\noindent 03.65.Sq ~ Semiclassical theories and applications. \\
\noindent 05.45.Mt ~ Semiclassical chaos (``quantum chaos'').

\newpage

\section{Introduction}

Correlations in the spectra of quantum systems with a chaotic
classical limit are expected to follow random matrix theory.
This is the content of the random matrix hypothesis
\cite{BGS84,Boh91}, which has found confirmation by
numerical investigations in many systems.
Theoretical support for this conjecture has been obtained
by semiclassical approximations based on periodic orbits
\cite{Ber85,BK96} as well as field theoretical methods \cite{AAA95}.
In particular, it has been shown by the semiclassical method
that the leading asymptotic behaviour of two-point correlation
functions for long-range correlations agrees with
results of random matrix theory. This result has been
obtained by using mean properties of periodic orbits
as expressed by the sum rule of Hannay and Ozorio de Almeida
\cite{HO84}.
It is expected that a proof of the random matrix hypothesis
would be possible if certain correlations between periodic
orbits were known \cite{ADDKKSS93}.

The semiclassical analysis is based on the trace formula
for the density of states \cite{Gut90}. This is a leading order
approximation, as $\hbar \rightarrow 0$, in terms
of the periodic orbits of the corresponding classical
system. For most systems this approximation is not
exact, and there are corrections in higher order of
$\hbar$. A particular important correction occurs
if the potential has discontinuities or singularities
which lead to wave diffraction. This is the case
in many standard examples of chaotic systems.
It leads to additional contributions to the trace formula
in terms of so-called creeping orbits or in terms of
trajectories that are closed or
connect two points in coordinate space
\cite{VWR94,PSc95,BW96,PSSU96}. In the
present article we investigate systems in which the
diffraction occurs at point-like objects, for example
at a delta-like singularity of the potential, at a 
magnetic flux line in a two-dimensional system,
or at a corner in a billiard system.
For these systems we examine whether diffractive 
corrections to the trace formula can have an
influence on spectral statistics that persists
in the semiclassical limit $\hbar \rightarrow 0$.
The analysis is done by deriving a classical sum
rule for the orbits that arise in the geometrical
theory of diffraction. With this input
the diffractive corrections to the diagonal
approximation for the spectral form factor are
calculated, and it is shown that these corrections
in general do not vanish as $\hbar \rightarrow 0$.

\section{Sum rules for transient orbits}

Semiclassical approximations for the Green function 
$G(\q_b,\q_a,E)$ of a quantum system involve classical
trajectories that go from $\q_a$ to $\q_b$ at energy $E$.
In general there is an infinite number of these trajectories.
Moreover, in a chaotic system the number of trajectories
that connect the two points in a time less than $T$
increases exponentially with $T$. In this section
we use the ergodic property of chaotic systems in
order to obtain a sum rule for orbits connecting
two points in coordinate space. We call these trajectories
transient orbits.

Consider a particle with energy $E$ that starts at time
$t=0$ at a point $\q_a$ in coordinate space. The classical
probability density for the particle to be found at time
$t=T$ at a point $\q_b$ is given by
\begin{equation} \label{prop}
P(\q_b;T,\q_a,E) = 
\frac{\int \! \dd^f p_a \; \delta(E - H(\q_a,\p_a)) \;
\delta(\q(T) - \q_b)} {\int \! \dd^f q_b \int \! \dd^f p_a \;
\delta(E - H(\q_a,\p_a)) \; \delta(\q(T) - \q_b)} \; ,
\end{equation}
where the function $\q(t)$ denotes the position of the particle
with initial position $\q_a$ and momentum $\p_a$ at $t=0$, and $f$
is the number of degrees of freedom of the system.
The integration in the numerator extends over all initial
momenta corresponding to energy $E$, and the denominator gives
the normalisation constant.
The probability density is only different from zero if there
are classical trajectories that connect the points $\q_a$
and $\q_b$ in time $T$. 

Let us assume that the particle moves in the field of a
vector and a scalar potential and that the Hamiltonian
is given by
\begin{equation}
H(\q,\p) = \frac{1}{2m} \left(\p - 
\frac{e}{c} \operatorname{A}(\q) \right)^2 + V(\q) \; .
\end{equation}
In the following we express the transition probability
density for this Hamiltonian in terms of classical
trajectories.

The numerator in (\ref{prop}) is evaluated by introducing
a local coordinate system in the vicinity of a trajectory
where one coordinate is along the trajectory and the
others are perpendicular to it,
\begin{equation}
\int \! \dd^f p_a \; \delta(E - H(\q_a,\p_a)) \;
\delta(\q(T) - \q_b)
= \sum_\gamma \frac{1}{v_a \, v_b} \, \left|
{\det}' \left( \frac{\partial \q_b}{\partial \p_a} \right)_\gamma 
\right|^{-1} \,
\delta(T_\gamma - T) \; .
\end{equation}
The sum is over all trajectories from $\q_a$ to $\q_b$
at energy $E$, $v_a$ and $v_b$ are the velocities at $\q_a$ 
and $\q_b$, respectively, $T_\gamma$ is the time along
the trajectory $\gamma$, and the prime at the determinant
denotes that the matrix of derivatives involves only the
coordinates orthogonal to the trajectory.

The denominator is evaluated by changing the integration
over momentum into one over velocity and evaluating it
in hyperspherical coordinates. The remaining integral
over coordinates gives one, and one obtains
\begin{equation}
\int \! \dd^f q_b \int \! \dd^f p_a \;
\delta(E - H(\q_a,\p_a)) \; \delta(\q(T) - \q_b)
= \frac{(m v_a)^{f-1}}{v_a} \, {\cal S}^{(f)} \; ,
\end{equation}
where ${\cal S}^{(f)} = 2 \, \pi^{f/2} / \Gamma(f/2)$
is the surface area of an $f$-dimensional hypersphere with
unit radius. Altogether the result is
\begin{equation} \label{traj}
P(\q_b;T,\q_a,E) = \sum_\gamma \frac{1}{v_b \, (m v_a)^{f-1} \,
{\cal S}^{(f)}} \, \left|
{\det}' \left( \frac{\partial \q_b}{\partial \p_a} \right)_\gamma
\right|^{-1} \,
\delta(T_\gamma - T) \; .
\end{equation}

In an ergodic system the property that typical trajectories
fill out phase space uniformly can be used to describe the
probability density for long times $T$. In order to apply
ergodicity we smooth the singular function $P$
over some range of the final coordinate
\begin{equation} \label{smooth}
\langle P(\q_b;T,\q_a,E) \rangle_{\varepsilon}
\equiv \int \! \dd^f q_b' \; P(\q_b';T,\q_a,E) \;
\delta_\varepsilon (\q_b -\q_b') \; ,
\end{equation}
where $\delta_\varepsilon$ is a smoothed delta-function
whose width is parameterised by $\varepsilon$, and which
is normalised to one 
\begin{equation}
\int_{V(\q) < E} \! \dd^f q \:
\delta_\varepsilon (\q - \q_b) = 1 \; .
\end{equation}
It is assumed that the width $\varepsilon$ is classically small
so that the potentials do not change appreciably within this
region. The integral in (\ref{smooth}) results in the
replacement of the second delta function in the numerator of
(\ref{prop}) by $\delta_\varepsilon$.

In order to describe the properties of the probability
density $P$ we employ the property of an ergodic
system that the time-average of a function along a 
trajectory is equal to its phase space average for almost
all trajectories
\begin{equation} \label{ergod}
\lim_{T \rightarrow \infty} \frac{1}{T} \; \int_{T_0}^{T} \! \dd t
\; f(\q(t),\p(t)) = \frac{\int \! \dd^f q \, \dd^f p \; 
\delta(E - H(\q,\p)) \; f(\q,\p)}{
\int \! \dd^f q \, \dd^f p \; \delta(E - H(\q,\p))} \; ,
\end{equation}
and we choose for $f$ the smoothed delta-function
$\delta_\varepsilon$. The phase space average over
this function is evaluated similarly as before and yields
\begin{equation} \label{aver}
 \frac{\int \! \dd^f q \, \dd^f p \; 
\delta(E - H(\q,\p)) \; \delta_\varepsilon(\q - \q_b)}{
\int \! \dd^f q \, \dd^f p \; \delta(E - H(\q,\p))}
= \frac{(m v_b)^{f-1} \, {\cal S}^{(f)}}{v_b \, \Sigma(E)}
+ {\cal O}(\varepsilon) \; ,
\end{equation}
where $\Sigma(E)$ is the volume of the energy surface
in phase space. After integrating (\ref{ergod}) with
(\ref{aver}) over the initial momenta and dividing by
the normalisation constant one obtains the following
result for the probability density
\begin{equation} \label{sumprop}
\lim_{T \rightarrow \infty}
\frac{1}{T} \; \int_{T_0}^{T} \! \dd t \;
\langle P(\q_b;t,\q_a,E) \rangle_{\varepsilon}
= \frac{(m v_b)^{f-1} \, {\cal S}^{(f)}}{v_b \, \Sigma(E)}
+ {\cal O}(\varepsilon) \; .
\end{equation}
This property of $P$ has the following
interpretation. When the first term on the right-hand side
is multiplied by a volume element $\dd^f q$ it is the
ratio between the volume of the part of the energy
surface in phase space corresponding the element
$\dd^f q$ around $\q_b$ and the volume of the total
energy surface. Thus the probability of a particle
to be found in a neighbourhood of $\q_b$ is equal
to the relative volume of the energy surface of this
neighbourhood.

Equation (\ref{sumprop}) gives an expression for the
probability density $P$ that is smoothed over an
$\varepsilon$-neighbourhood of the final point in an
ergodic system. Using (\ref{traj}), it is a sum rule
for classical trajectories that start at $\q_a$ and
visit the $\varepsilon$-neighbourhood of $\q_b$.
In order to obtain a sum rule for trajectories that
hit $\q_b$ one has to take the limit $\varepsilon
\rightarrow 0$. Here one faces the problem that
one would like to interchange the two limits
$\varepsilon \rightarrow 0$ and $T \rightarrow \infty$,
a problem that occurs also in the derivation of
sum rules for periodic orbits \cite{HO84}.

One can argue that the two limits can be interchanged
if the classical trajectories do not have conjugate
points. Then the terms 
$[{\det}' (\partial \q_b/\partial\p_a)]^{-1}$
decrease exponentially with the transition
time $T_\gamma$. This means that the contribution
of a single trajectory to the probability density
is exponentially small for long times. The integral
over the probability density $\int^T_{T_0} \dd t 
\, P(\q_b;t,\q_a,E)$ is a discontinuous function
of the final coordinate $\q_b$, but its variation
inside the $\varepsilon$-environment of $\q_b$
becomes very small for long times $T$, and the
two limits $\varepsilon \rightarrow 0$ and
$T \rightarrow \infty$ can be interchanged.

This argumentation does not apply, however, if
there are caustics of the classical motion inside
the $\varepsilon$-neighbourhood. On a caustic
the term $[{\det}' (\partial \q_b/\partial\p_a)]^{-1}$
is divergent. This divergence is integrable so that
the smoothed version of the sum rule (\ref{sumprop})
is still valid. It might be possible that in
the generic situation where the point $\q_b$ itself
does not lie on a caustic, the two limits can still
be interchanged. We point out, however, that for
applications in semiclassical approximations
quantum mechanics provides a natural smoothing of
the probability density, so that only a smoothed
version of the sum rule is required.

In cases where the two limits can be interchanged
we obtain the following sum rule for trajectories
from $\q_a$ to $\q_b$ at energy $E$
\begin{equation}
\lim_{T \rightarrow \infty}
\frac{1}{T} \sum_{T_\gamma < T} 
\frac{1}{v_b \, (m v_a)^{f-1} \, {\cal S}^{(f)}} \, \left|
{\det}' \left( \frac{\partial \q_b}{\partial \p_a} \right)_\gamma
\right|^{-1} 
= \frac{(m v_b)^{f-1} \, {\cal S}^{(f)}}{v_b \, \Sigma(E)} \; ,
\end{equation}
and the differentiated version is
\begin{equation} \label{sumt}
\sum_\gamma \frac{1}{v_b \, (m v_a)^{f-1} \, {\cal S}^{(f)}} \, \left|
{\det}' \left( \frac{\partial \q_b}{\partial \p_a} \right)_\gamma 
\right|^{-1} \, \delta(T- T_\gamma)
\approx \frac{(m v_b)^{f-1} \, {\cal S}^{(f)}}{v_b \, \Sigma(E)} \, ,
\qquad T \rightarrow \infty \; ,
\end{equation}
where it is implied that the left-hand side has to be smoothed
over some small time-interval $\Delta T$ in order to obtain a
smooth function.

In the following we give several variants of the
sum rule that can be obtained in an analogous way.
\begin{itemize}
\item For chaotic area-preserving maps on an $(2f)$-dimensional
unit torus, $(\q_{n+1},\p_{n+1}) = h(\q_n,\p_n)$, the
corresponding result is
\begin{equation} \label{summap}
\sum_{(\q_n,\q_0)=(\q_b,\q_a)} \left| \det \left( \frac{
\partial \q_n}{\partial \p_0} \right) \right|^{-1} \approx 1 \; , 
\qquad n \rightarrow \infty \; ,
\end{equation}
where the sum extends over all points $(q_0,p_0)$ for
which $q_0=q_a$ and $q_n=q_b$. In appendix \ref{cats} it is
shown that this sum rule is exact for cat maps for all $n>0$.
\item For billiard systems the sum rule (\ref{sumt}) can be
expressed in a geometrical form. In two dimensions the result is
\begin{equation} \label{sumbill}
\frac{1}{2 \pi} \sum_\gamma \frac{1}{|M_{12}|}
\delta(L - L_\gamma) \approx \frac{1}{A} \; , 
\qquad L \rightarrow \infty \; ,
\end{equation}
where $A$ is the area of the billiard, $L_\gamma$ is the
geometrical length of a trajectory and
$M_{12} = p_a \, (\partial q_b^\perp / \partial p_a^\perp)$
is an element of the stability matrix, scaled by $p_a$ in
order to be energy independent. Eq.~(\ref{sumbill}) expresses
that a particle is equally likely to be found anywhere in the
billiard if it travels a sufficiently long distance $L$.
Consequently, the probability density that it is found at some
point $\q_b$ is equal to $A^{-1}$.
\item For applications in the next section one requires
sum rules for a subset of transient orbits for which
the angular orientations of the initial and the final
velocities, $\pphi_a$ and $\pphi_b$, are fixed,
\begin{equation} \label{sumpart}
\sum_{\pphi_a,\pphi_b \, \text{fixed}} 
\left| {\det}' \left( \frac{\partial \q_b}{\partial \p_a} 
\right)_\gamma \right|^{-1} \,
\delta(T- T_\gamma) \approx \frac{(m v_a)^{f-1} \, (m v_b)^{f-1}
}{\Sigma(E)} \, \dd \Omega_a \, \dd \Omega_b \; ,
\qquad T \rightarrow \infty \; .
\end{equation}
Here the sum extends over all transient trajectories for which the
initial and final angular directions lie in solid angle elements
$\dd \Omega_a$ and $\dd \Omega_b$, respectively, around the
directions $\pphi_a$ and $\pphi_b$.
\item In integrable systems a similar sum rule can be obtained,
if the ergodic average is performed over an invariant torus instead
of the energy surface.
\begin{align} \label{sumint}
& \sum_{\pphi_a \, \text{fixed}} \frac{1}{v_b \, (m v_a)^{f-1} \, 
{\cal S}^{(f)}} \,
\left| {\det}' \left( \frac{\partial \q_b}{\partial \p_a}
\right)_\gamma \right|^{-1} \,
\delta(T- T_\gamma) \notag \\ \approx &
\frac{1}{(2 \pi)^f \, {\cal S}^{(f)}} \, 
\sum_{\I(\q_b,\p_b) = \I_a}
\left| \det \left( \frac{\partial \I(\q_b,\p_b)}{\partial \p_b}
\right) \right|^{-1} \, \dd \Omega_a
\, , \quad T \rightarrow \infty \; ,
\end{align}
where $\I$ are the action variables of the system. This equation
has the following interpretation. The left-hand side is a sum
over all trajectories for which the initial angular direction
lies in a solid angle element $\dd \Omega_a$ around the direction
$\pphi_a$. This direction $\pphi_a$, the energy $E$ and the
initial position $\q_a$ determine the action variable
$\I_a=\I(\q_a,\p_a)$ and thus the torus on which the motion
occurs. The right-hand side is a summation over all final
momenta $\p_b$ for which the points $(\q_b,\p_b)$ lie on
this torus.

In order to obtain a sum rule for all transient trajectories
one has to integrate (\ref{sumint}) over all initial directions.
In contrast to the chaotic case this is not always possible,
since the integral can be divergent due to caustics. In cases
where it is possible one has
\begin{align} \label{sumint2}
& \sum_{\gamma} \frac{1}{v_b \, (m v_a)^{f-1} \,
{\cal S}^{(f)}} \, \left|
{\det}' \left( \frac{\partial \q_b}{\partial \p_a}
\right)_\gamma \right|^{-1} \,
\delta(T- T_\gamma) \notag \\ \approx & 
\frac{1}{(2 \pi)^f \, {\cal S}^{(f)}} \, 
\int \! \dd \Omega_a \, \sum_{\I(\q_b,\p_b) = \I_a}
\left| \det \left( \frac{\partial \I(\q_b,\p_b)}{\partial \p_b}
\right) \right|^{-1} \, , \quad T \rightarrow \infty \; .
\end{align}
This sum rule can be verified, for example, 
for a $f$-dimensional rectangular billiard with side
lengths $a_i$, $i=1,\dots,f$. There the
number of transient trajectories with length in an 
interval $\dd L$ around $L$ is given asymptotically for
large $L$ by ${\cal S}^{(f)} \, L^{f-1} \, \dd  L/V$
where $V=\prod_{i=1}^n a_i$ is the volume of the billiard,
and $L=v_a \, T$. The absolute value of the determinant on
the left-hand-side of
(\ref{sumint2}) is $(L/(m v_a))^{f-1}$, the actions
are $I_i=a_i \, |p_i|/\pi$, and the sum on the
right-hand-side is $(2 \pi)^f/V$. Altogether one obtains
on both sides $V^{-1}$. Incidentally
the sum rule for chaotic billiard systems yields the same
result in this case. The reason is that for
rectangular billiards the determinant on the right-hand
side of (\ref{sumint2}) is the same for all tori.
\end{itemize}

The sum rule (\ref{sumt}) contains implicit information
about the number of transient orbits. Let us assume that
due to the exponential sensitivity of trajectories on
initial conditions one has
$\langle |{\det}' (\partial \q_b / \partial \p_a)|^{-1} 
\rangle \sim c' \, \exp(-h T)$ where $c'$ and $h$ are constants,
and the averaging is performed over trajectories with
$T_\gamma \approx T$. With this assumption one obtains
the following asymptotic law for the number of transient orbits,
\begin{equation} \label{number}
\rho(T) = \frac{\dd {\cal N}(T)}{\dd T} 
\sim c \, \exp(h T) \; , \qquad T \rightarrow \infty \; ,
\end{equation}
where ${\cal N}(T)$ is the number of transient trajectories
with $T_\gamma < T$, $\rho(T)$ the density of these orbits, and
$c$ is a constant that is determined by $c'$. For Riemannian
manifolds with no conjugate points one can show that 
$\lim_{T \rightarrow \infty} T^{-1} \log {\cal N}(T) = h$
is the topological entropy of the system \cite{BP97}. 
Eq.~(\ref{number}) with topological entropy $h$
would imply that the number of transient orbits grows 
by an order of $T$ stronger than the number of periodic
orbits for which $\rho_{po}(T) \sim \exp(h T)/T$.
One arrives at a similar conclusion for the number of transient
orbits if one considers systems in which a code for these
orbits exists \cite{BW96,RVW96,BD97}. For cat maps the law (\ref{number})
is proved in appendix \ref{cats}.

\section{Influence of diffractive orbits on spectral statistics}
\label{secstat}

Diffraction of quantum wave functions on singularities
of the potential leads to corrections in semiclassical
expansions in terms of classical trajectories.
In approximations for the density of states there
are additional terms besides periodic orbits. These
are expressed in terms of diffractive orbits.
We concentrate on cases where the source of
diffraction is point-like for which some examples
are given below. In these cases the diffractive
orbits are trajectories that start from and return
to the source of diffraction $n$ times, where $n$
is an arbitrary positive integer. They are composed
of a sequence of arbitrary $n$ transient orbits
of the last section for which the initial and
the final points of the trajectory are identical
and are located at the source of diffraction.
Alternatively, they can be considered as closed
trajectories that are scattered $n$ times on the
source of diffraction.

Within the geometrical theory of diffraction (GTD)
the contribution of all diffractive orbits with $n$
scattering events to the density of states is
given by \cite{VWR94,PSc95,BW96}
\begin{align} \label{gtd}
d^{(n)}(E) & = \frac{1}{\pi \, n} \left( 
\frac{\hbar^2}{2 m} \right)^n \frac{\dd}{\dd E} \Im \left[
\sum_{\gamma_1} \dots \sum_{\gamma_n} {\cal G}_{\gamma_1}(E) 
{\cal D}(\pphi_{a,\gamma_{2}},\pphi_{b,\gamma_1}) 
{\cal G}_{\gamma_2}(E) 
{\cal D}(\pphi_{a,\gamma_{3}},\pphi_{b,\gamma_2}) \right.
\notag \\ & \left. \phantom{\sum_{\gamma_n}} \cdots
{\cal G}_{\gamma_n}(E)
{\cal D}(\pphi_{a,\gamma_{1}},\pphi_{b,\gamma_n})
\right] \; ,
\end{align}
where
\begin{equation} \label{green}
{\cal G}_{\gamma_i}(E) =
\frac{1}{i \hbar (2 \pi i \hbar)^{(f-1)/2} } \,
\sqrt{\frac{1}{v_a \, v_b} \, 
\left| {\det}' \left( 
\frac{\partial \q_b}{\partial \p_a} \right)_{\gamma_i} \right|^{-1}}
\, \exp \left\{ \frac{i}{\hbar} S_{\gamma_i}(E)
- i \frac{\pi}{2} \nu_{\gamma_i} \right\} \; .
\end{equation}
The expression (\ref{green}) is the semiclassical contribution
of a transient orbit $\gamma_i$ to the Green function, and
$S_{\gamma_i}$ is its action and $\nu_{\gamma_i}$ the number
of conjugate points along it. Furthermore, $v_a$ and $v_b$
are the initial and final velocities. They are identical
since the initial and final point of the trajectory are
identical, and we write them in the following without subscript.

A diffractive orbit is composed
of $n$ transient orbits, and it is labelled by a set
of $n$ indices $\gamma_1, \dots , \gamma_n$. It is 
customary to consider cyclic permutations of these
indices to correspond to the same diffractive orbit.
The contribution of a diffractive orbit
to the density of states in (\ref{gtd}) can be interpreted in
the following way. In the GTD approximation
the diffraction is treated as a local process that
occurs at the source of diffraction. The diffraction
coefficient ${\cal D}(\pphi_{a,\gamma_{i+1}},\pphi_{b,\gamma_i})$
contains the amplitude and phase for the scattering from
the incoming direction of the transient orbits $\gamma_i$
into the outgoing direction of the transient orbit $\gamma_{i+1}$.
In some sense, the diffractive trajectories can be considered
as generalised periodic orbits which are closed in momentum space
by the scattering at the singularity of the potential.

Some examples for the diffraction coefficients are
\begin{itemize}
\item Diffraction on a corner with angle $\theta$ in a 
two-dimensional billiard system \cite{Kel62,Jam76}:
\begin{equation}
{\cal D}(\phi_a,\phi_b) = \frac{2}{N} \,
\sin\frac{\pi}{N} \left[ 
\left(\cos\frac{\pi}{N}
-\cos\frac{\phi_a+\phi_b}{N} \right)^{-1} - 
\left(\cos\frac{\pi}{N}
-\cos\frac{\phi_a-\phi_b}{N} \right)^{-1}
\right] \; ,
\end{equation}
where $N=\theta/\pi$, and $\phi_a$ and $\phi_b$ are the angles
of the outgoing and incoming trajectory, measured with respect
to one side of the corner such that $\phi_a,\,\phi_b \in [0,\theta]$.
\item Diffraction on a flux line in a two-dimensional
system \cite{Sie99a}:
\begin{equation}
{\cal D}(\phi_a,\phi_b) = \frac{2 \sin(\alpha \pi)}{
\cos \left(\frac{\phi_a - \phi_b}{2} \right) }
\exp\left\{ i \frac{\phi_a - \phi_b}{2} \right\} \; ,
\end{equation}
where $\alpha$ is the flux parameter and $\phi_a$ and $\phi_b$ are
the directions of the outgoing and incoming trajectory.
\item Diffraction on a delta-like singularity of the potential
in two dimensions \cite{AGHH88,ES96}:
\begin{equation}
{\cal D} = \frac{2 \pi}{i \frac{\pi}{2} - \gamma - 
\log\left(\frac{k a}{2}\right)} \; ,
\end{equation}
where $k=\sqrt{2 m E}/\hbar$ and $a$ is a parameter characterising
the strength of the potential, and $\gamma$ is Euler's constant.
\item Diffraction on a delta-like singularity of the potential
in three dimensions \cite{AGHH88,ES96}: 
\begin{equation}
{\cal D} = \frac{4 \pi a }{1 + i k a} \; .
\end{equation}
\end{itemize}

In the first two examples the diffraction coefficient is energy
independent but it depends on the incoming
and outgoing directions of the trajectories. There are directions
in which the diffraction coefficient diverges. There the GTD
approximation breaks down and has to be replaced by a uniform
approximation \cite{SPS97,Sie99a}. In the last two cases, the
diffraction coefficient has no angular dependence, but depends
on energy. It corresponds to pure $s$-wave scattering, and
the GTD approximation is valid for all angular directions.

In the following we consider the influence of diffractive
orbits on semiclassical approximations for the spectral
form factor which is defined by
\begin{equation}
K(\tau) = \int_{-\infty}^{\infty} \! \frac{\dd \eta}{\bar{d}(E)} \;
\left\langle d_{\text{osc}}\left( E + \frac{\eta}{2} \right)
             d_{\text{osc}}\left( E - \frac{\eta}{2} \right)
\right\rangle_E
\; \exp\left( 2 \pi i \eta \tau \bar{d}(E) \right) \; ,
\end{equation}
where $\langle \dots \rangle_E$ denotes an average over an energy
interval that is small in comparison with $E$ but contains many
energy levels. If the oscillatory part of the density of states
is semiclassically approximated by classical trajectories
in the form 
\begin{equation} \label{density}
d_{\text{osc}}(E) \approx \sum_\gamma A_\gamma
\exp\left( \frac{i}{\hbar} S_\gamma(E) \right) \; ,
\end{equation}
then the spectral form factor is expressed by a double sum over
trajectories. We consider here only the diagonal 
approximation to this double sum that describes the form
factor for small values of $\tau$. In leading semiclassical order
\begin{equation}
K_d(\tau) = \frac{2 \pi \hbar}{\bar{d}(E)} \sum_\gamma \, g_\gamma \,
|\bar{A}_\gamma|^2 \, \delta(T-T_\gamma) \; ,
\end{equation}
where $T=2 \pi \hbar \bar{d}(E) \tau$ and $g_\gamma$ is the number
of terms in the sum for which the actions $S_\gamma(E)$ are identical.
$\bar{A}_\gamma$ is the average over all $g_\gamma$ amplitudes of
the orbits which have the same action as the orbit $\gamma$ in
case that these amplitudes are different.

Let us first consider the case of single-diffractive orbits $n=1$. Then
\begin{equation}
A_\gamma = \frac{T_\gamma \, {\cal D}(\pphi_{a,\gamma},\pphi_{b,\gamma})
}{4 \pi m v (2 \pi \hbar)^{(f-1)/2}}
\sqrt{\left| {\det}' \left( 
\frac{\partial \q_b}{\partial \p_a} \right)_\gamma \right|^{-1}} \; ,
\end{equation}
and one obtains for the contribution of diffractive orbits with
one scattering event to the diagonal approximation of the form
factor
\begin{equation}
K^{(1)}_d(\tau) = \frac{2 \pi \hbar (2/\beta)}{\bar{d}(E)} \sum_{\gamma} 
\frac{T_\gamma^2 \, |{\cal D}(\pphi_{a,\gamma},\pphi_{b,\gamma})|^2
}{(4 \pi m v)^2 \, (2 \pi \hbar)^{f-1}} \, \left| {\det}' \left( 
\frac{\partial \q_b}{\partial \p_a} \right)_\gamma \right|^{-1} \,
\delta(T - T_{\gamma}) \; .
\end{equation}
The degeneracy of the actions is in this case $g=2/\beta$
(except for an exponentially small fraction of the orbits for large $T$)
where $\beta$ is an integer denoting the symmetry class of the
system. $\beta=1$ for systems in which the only symmetry is
an anti-unitary symmetry, and $\beta=2$ for systems without
any symmetry. Applying the sum rule (\ref{sumpart}) and the
leading asymptotic approximation for the mean density of states
$\bar{d}(E) \sim \Sigma(E)/(2 \pi \hbar)^f$ one obtains
\begin{equation}
\dd^2 K^{(1)}_d(\tau) = \frac{1}{8 \beta \pi^2} \,
\left( \frac{m v}{2 \pi \hbar} \right)^{2 f-4} \, 
|{\cal D}(\pphi_a,\pphi_b)|^2 \, \tau^2 \, \dd \Omega_a \dd \Omega_b
\end{equation}
for the partial contribution to $K^{(1)}_d(\tau)$ from diffractive
orbits with fixed initial and final directions.

For the examples given above this result has the following
implications. For the corner diffraction and diffraction
on a flux line the prefactor of $\tau^2$ is independent
of $\hbar$. It implies that diffraction has an influence
on statistical properties of energy levels in the
semiclassical limit $\hbar \rightarrow 0$. Although
the contributions of diffractive orbits to the density
of states are by an order $\sqrt{\hbar}$ smaller than
those of periodic orbits, they still give a finite
contribution to the diagonal approximation of the
spectral form factor as $\hbar \rightarrow 0$.
The complete contribution to $K_d^{(1)}$ that is
obtained by integrating over all angular directions
$\pphi_a$ and $\pphi_b$ requires the use of uniform
approximations, since the GTD approximation yields a
divergent result. However, since all contributions to
$K_d^{(1)}$ are positive there is a non-vanishing
contribution of order $\tau^2$ to the diagonal approximation
of the form factor in the semiclassical limit.
For the examples of diffraction on delta-singularities
the diffraction is isotropic and the integration over
the angular directions can be performed as is done
below. Here the prefactor of $\tau^2$ is
energy dependent, and there is a significant difference
between the two- and three-dimensional result. In two
dimensions the prefactor vanishes in the semiclassical
limit, although very slowly like $(\log \hbar)^{-2}$, whereas
in three dimensions the prefactor approaches a constant.

The contributions of diffractive orbits with $n$ scattering
events to $K_d(\tau)$ are obtained in a similar way. They
are composed of $n$ transient orbits and it is sufficient
to consider only cases in which all these $n$ orbits
are different, since the relative number of the other cases
is exponentially suppressed for large times $T$. The
degeneracy of the actions of diffractive orbits is
then $g=(2/\beta)^n \, n!$. The index $\gamma$ in
(\ref{density}) is now a multiple index
$\gamma = (\gamma_1,\dots,\gamma_n)$ and
\begin{equation}
\bar{A}_\gamma = \frac{T_\gamma}{2 \pi \hbar n} \, 
\left( \frac{\hbar}{2 m v (2 \pi \hbar)^{(f-1)/2}} \right)^n
\overline{\cal D}^{(n)}(\pphi_{a,\gamma_1},\pphi_{b,\gamma_1},
\dots,\pphi_{a,\gamma_n},\pphi_{b,\gamma_n})
\prod_{i=1}^n
\sqrt{\left| {\det}' \left( 
\frac{\partial \q_b}{\partial \p_a} \right)_{\gamma_i} \right|^{-1}} \; ,
\end{equation}
where $T_\gamma = \sum_{i=1}^n T_{\gamma_i}$, and
\begin{equation}
\overline{\cal D}^{(n)}(\pphi_{a,\gamma_1},\pphi_{b,\gamma_1},
\dots,\pphi_{a,\gamma_n},\pphi_{b,\gamma_n})
= \left\langle \prod_{i=1}^n 
{\cal D}(\pphi_{a,\gamma_{i+1}},\pphi_{b,\gamma_i}) \right\rangle
\end{equation}
is the average over all $n!$ permutations of the $\gamma_i$.
For simplicity of notation we abbreviate the argument of
$\overline{\cal D}^{(n)}$ by an index $\gamma$. Using the relation
\begin{equation}
\delta(T - T_\gamma) = \int_0^\infty \! \dd T_1 \cdots \dd T_n \;
\left[ \prod_{i=1}^n \delta(T_i - T_{\gamma_i}) \right] \,
\delta( T - \sum_{i=1}^{n} T_i) \; ,
\end{equation}
one can write the contribution of $n$-fold diffractive orbits
to the diagonal approximation of the form factor in the form
\begin{align}
K_d^{(n)}(\tau) & = \frac{T^2 (2/\beta)^n n!}{2 \pi \hbar n^2
\bar{d}(E)} \, \int_0^\infty \! \dd T_1 \cdots \dd T_n \;
\delta( T - \sum_{i=1}^{n} T_i) \, \times \notag \\ & \qquad
\sum_\gamma |\overline{\cal D}_\gamma^{(n)}|^2
\prod_{i=1}^n \left[
\frac{1}{(4 \pi m v_a)^2 (2 \pi \hbar)^{f-3}} \,
\left| {\det}' \left( 
\frac{\partial \q_b}{\partial \p_a} \right)_{\gamma_i} \right|^{-1} \,
\delta(T_i - T_{\gamma_i}) \right] \; .
\end{align}
After using sum rule (\ref{sumpart}) $n$ times one is left with the
integrals over the $T_i$ which yield
\begin{equation}
\int_0^\infty \! \dd T_1 \cdots \dd T_n \;
\delta( T - \sum_{i=1}^{n} T_i) = \frac{T^{n-1}}{(n-1)!} \; ,
\end{equation}
which can be shown by induction, and the final result is
\begin{equation}
\dd^{2 n} K^{(n)}_d(\tau) = \frac{\tau^{n+1}}{n} \,
\left( \frac{1}{8 \beta \pi^2} \,
\left( \frac{m v}{2 \pi \hbar} \right)^{2 f-4} \, \right)^n 
\, |\overline{\cal D}^{(n)}(\pphi_{a,1},\dots,\pphi_{b,n})|^2 \,
\dd \Omega_{a,1} \cdots \dd \Omega_{b,n} \; .
\end{equation}
The interpretation is similar as before. The $n$-fold diffractive
orbits contribute to $K_d(\tau)$ in order $\tau^{n+1}$. For
corner diffraction and diffraction on a flux line the
prefactor of $\tau^{n+1}$ is independent of $\hbar$ and persists
in the semiclassical limit. The remarkable point is that
the order in $\hbar$ of the contributions of these orbits
to the density of states can be arbitrarily large, their
amplitude is by an order $\hbar^{n/2}$ smaller than those
of periodic orbits, but they still give a finite contribution
to the form factor in the semiclassical limit! For the
diffraction on delta-singularities of the potential,
the prefactor of $\tau^{n+1}$ again vanishes slowly in
the two-dimensional case and approaches a constant in
three dimensions as $\hbar \rightarrow 0$.

For the isotropic case where the diffraction coefficient
is angular independent, the integration over the angular
directions can be performed, and the contributions of
all diffractive orbits can be summed. Adding also the
contribution of periodic orbits, the complete expression
for the diagonal form factor is
\begin{equation}
K_d(\tau) = \frac{2}{\beta} \tau + \sum_{n=1}^\infty 
\frac{\tau^{n+1}}{n} 
{\cal C}^n = \frac{2}{\beta} \tau - \tau \, \log (1-{\cal C} \tau)
\; , \qquad \qquad \tau < {\cal C}^{-1} \; ,
\end{equation}
where 
\begin{equation}
{\cal C} = \frac{|{\cal D}|^2 \, [{\cal S}^{(f)}]^2}{8 \beta \pi^2} \,
\left( \frac{m v}{2 \pi \hbar} \right)^{2 f-4} \, .
\end{equation}
For an $s$-wave scatterer in a three-dimensional billiard
system the constant approaches the value ${\cal C}=8$ 
in the limit 
$\hbar \rightarrow 0$. 

The results
for the contributions of diffractive orbits to the
spectral form factor
can also be applied to a rectangular billiard with
an $s$-wave scatterer, since the same sum-rule
applies in this case. The form factor for this 
system has been studied in \cite{DV98}.

\section{Conclusions}

The main result of this article is that diffraction can
have an influence on spectral statistics in the
semiclassical limit. The semiclassical treatment of
diffraction leads to corrections to the diagonal
approximation for the spectral form factor $K(\tau)$
that do not vanish in the semiclassical limit
$\hbar \rightarrow 0$. This is
in contrast, for example, to corrections due to bouncing
ball orbits, which vanish if the semiclassical limit
is performed while keeping the argument of $K(\tau)$ fixed.

Although the corrections to the form factor do not agree
with those expected from random matrix theory,
the present results do not show that diffraction leads
to a deviation from random matrix statistics, since
off-diagonal contributions to the form factor have 
been neglected. They show, however, that a semiclassical
proof of the random matrix hypothesis cannot be based
on periodic orbits alone if diffraction occurs.
Consider, for example, chaotic billiard systems with
flux lines. These systems are considered to be standard
examples in which spectral correlations are expected to
follow the statistics of the Gaussian unitary ensemble
(GUE). The diagonal approximation for the form factor
in terms of periodic orbits already yields the correct 
linear GUE form factor up to the Heisenberg time $\tau=1$
\cite{Ber85}. 
Moreover, it can be shown that in certain ensemble
averages the off-diagonal contributions of periodic
orbits vanish \cite{FK98}. The present results show,
however, that diffraction on the flux line leads
to additional semiclassical contributions to the
diagonal form factor in all higher orders of $\tau$.
If the spectral statistics follow indeed random
matrix theory, then these additional contributions
have to be cancelled by off-diagonal terms involving
diffractive orbits. Such a cancellation would require
specific correlations between classical trajectories
that start from and return to one point in coordinate
space, or between these trajectories and 
periodic orbits, analogous to the action correlations
of periodic orbits \cite{ADDKKSS93}.

For another example let us assume that the spectral
correlations in a particular three-dimensional chaotic
system are described correctly by random matrix 
theory in the semiclassical limit. Then it is generally
expected that this property is not changed, if an
$s$-wave scatterer is added to the system. As is
shown in the present article, however, the
diagonal approximation for $K(\tau)$
is modified also in this case. This would imply a
deviation from random matrix results, if this is not
corrected by the off-diagonal terms. Nevertheless,
it shows that a single $s$-wave scatterer {\it can}
lead to deviations from random matrix statistics.

The analysis in this article is based on the geometrical
theory of diffraction. For corner diffraction or
diffraction on a flux line this is not sufficient
for calculating the complete diagonal approximation
of the form factor. It would require the application
of uniform approximations. Since the semiclassical
weight of diffractive orbits is larger in
the uniform regime, one can expect an even stronger
total influence of diffraction on the spectral form
factor in these cases. The main remaining question
is, whether deviations from random matrix statistics
can be observed in these systems.
\bigskip \bigskip

\noindent
{\large \bf Acknowledgement}
\bigskip

\noindent
It is a pleasure to thank 
R.\ Artuso, 
E.\ Bogomolny, 
S.\ Fishman, 
E.\ Gutkin, 
J.\ Keating, and
U.\ Smilansky
for informative and helpful discussions on this subject.

\appendix

\section{Cat maps}
\label{cats}

In this appendix we give a simple example for the sum rule
for transient orbits. Cat maps are linear, area-preserving,
hyperbolic maps on the unit 2-torus
\begin{equation}
\begin{pmatrix} q_{n+1} \\ p_{n+1} \end{pmatrix} = M \, 
\begin{pmatrix} q_n     \\ p_n     \end{pmatrix} \; \mod 1 \; ,
\end{equation}
where $\det M = 1$ due to area preservation, $|\Tr M| > 2$
due to hyperbolicity, and the elements $M_{ij}$ of the matrix
$M$ are integers for continuity.

The sum rule (\ref{summap}) has the form 
\begin{equation} \label{sumcat}
\sum_{(q_n,q_0)=(q_b,q_a)} \frac{1}{|(M^n)_{12}|} = 1 \; ,
\end{equation}
where the sum extends over all points $(q_0,p_0)$ for
which $q_0=q_a$ and $q_n=q_b$. The matrix element $(M^n)_{12}$
is the same for all points and can be taken in front of the
sum. The number of points over which the sum extends 
is given by the number of solutions of the equation
\begin{equation}
q_b = (M^n)_{11} \, q_a + (M^n)_{12} \, p_a \; \mod 1 \; .
\end{equation}
Since $p_a$ varies in the interval $[0,1)$ there are
exactly $|(M^n)_{12}|$ solutions of this equation.
This shows that the sum rule (\ref{sumcat}) is exact
for all $n>0$.

Let $|\Tr M| = 2 \cosh(\lambda)$ where $\lambda>0$ is the
Lyapunov exponent of the map, and $\sigma =
\operatorname{sign}(\Tr M)$. Then
\begin{equation}
M^n = \frac{1}{u_1 s_2 - u_2 s_1} \,
\begin{pmatrix} u_1 & s_1 \\ u_2 & s_2 \end{pmatrix}
\begin{pmatrix} \sigma^n e^{n \lambda} & 0  \\
                0 & \sigma^n e^{-n \lambda} \end{pmatrix}
\begin{pmatrix} s_2 & -s_1 \\ -u_2 & u_1 \end{pmatrix} \; ,
\end{equation}
where $(u_1,u_2)$ and $(s_1,s_2)$ are the components of
the unstable and stable eigenvectors of $M$,
respectively. One obtains $(M^n)_{12} = - \sigma^n
u_1 s_1 (e^{n \lambda} + e^{-n \lambda})/(u_1 s_2 - u_2 s_1)$
and it follows that the number of points contributing to
the sum rule is given by
\begin{equation}
\rho(n) \sim \frac{|u_1 \, s_1|}{|u_1 s_2 - u_2 s_1|}
\, e^{n \lambda} \; , \qquad
n \rightarrow \infty \, .
\end{equation}
This is of the form (\ref{number}) and it shows, moreover,
that the constant $c$ in (\ref{number}) is not universal. 
For cat maps it is determined by the directions of the
eigenvectors of $M$.

\end{document}